\documentclass[twoside,10pt,a4paper]{newFNLstyle}
\usepackage{graphics}
\usepackage{cite}

\usepackage[]{graphicx}
\usepackage{amsmath}

\begin{document}

\volnumpagesyear{0}{0}{000--000}{2001}
\dates{received date}{revised date}{accepted date}

\title{DISORDER AND FLUCTUATIONS IN NONLINEAR EXCITATIONS IN DNA}

\authorsone{SARA CUENDA AND ANGEL S\'ANCHEZ$^{\dag}$}
\affiliationone{Grupo Interdisciplinar de Sistemas Complejos (GISC) and Departamento
de Matem\'aticas, Universidad Carlos III de Madrid, 28911 Legan\'es, Madrid, Spain\\
$^{\dag}$Instituto de Biocomputaci\'on y F\'\i sica de Sistemas Complejos (BIFI), 
Universidad de Zaragoza, 50009 Zaragoza, Spain}

\maketitle

\markboth{Disorder and fluctuations in nonlinear excitations in DNA}{Cuenda, S\'anchez}

\pagestyle{myheadings}



\begin{abstract}
We study the effects of the sequence on the propagation of nonlinear excitations
in simple models of DNA, and how those effects are modified by noise.
Starting from previous results on soliton dynamics on 
lattices defined by aperiodic potentials, [F.\ Dom\'\i nguez-Adame {\em et al.},
Phys.\ Rev.\ E {\bf 52}, 2183 (1995)], we analyze the behavior of lattices built from 
real DNA sequences obtained from human genome data. 
We confirm the existence of threshold forces, already found in Fibonacci sequences,
and of stop positions highly dependent on the specific sequence.
Another relevant conclusion is that the effective potential, a collective coordinate
formalism introduced by 
Salerno and Kivshar [Phys.\ Lett.\ A {\bf 193}, 263 (1994)] is a useful tool to
identify key regions that control the behaviour of a larger sequence.
We then study how the fluctuations can assist the propagation process by
helping the excitations to escape the stop positions. Our conclusions point out
to improvements of the model which look promising to describe mechanical 
denaturation of DNA.
Finally,
we also consider how randomly distributed energy focus on the chain as a function
of the sequence. 
\end{abstract}

\section{INTRODUCTION}

After a quiet, 30 year revolution in the way we look at natural phenomena, 
nonlinear science is nowadays a well established, respected body of knowledge 
with ubiquitous applications \cite{Scott}. Restricted at first to the 
realm of physics, nonlinear models are becoming a versatile tool in other,
unrelated fields. Such is the case of biology, where the nonlinear viewpoint
was successfully introduced more than 20 years ago \cite{Davydov} 
(see also Ref.\ \cite{Davydov2} for references). Ever since the early 
years, one of the subjects where the use of nonlinear models has been 
more productive is the modelling of DNA physics \cite{Yak1,Gaeta1,Yak2,Gaeta2}.
Indeed, the first paper on this issue dates back to 1980, when 
Englander and coworkers published a report entitled ``Nature of the open 
state in long polynucleotide double helices: possibility of soliton 
excitations'' \cite{Englander}. Subsequently, a large number of researchers
contributed to this field, which has by now reached maturity; the 
reader is referred to Ref.\ \cite {Yak2} for a historical account and 
an extensive summary of the available results.

Models such as that proposed by Englander and coworkers (hereafter, the 
Englander model) \cite{Englander} as well as other simple models proposed
by Peyrard and Bishop \cite{PB} (see also the generalizations and
improvements proposed in Refs.\ \cite{DPB,DP})
or Causo {\em et al.} \cite{Causo} have
demonstrated that, in spite of their simplicity, they can accurately 
(sometimes even quantitatively) capture the phenomenology experimentally
observed in DNA (see, e.g., Ref.\ \cite{Campa}). Most of the
research done in the framework of these models refers to homopolymers, 
i.e., homogeneous DNA molecules. In additions, a great deal of the 
experimental tests were carried out from a more thermodynamical or 
statistical mechanics viewpoint, focusing, e.g., on the characterization
of the thermal denaturation transition (the phase transition that takes 
place at temperatures around 70$^{\circ}$ C when the two strands of the
DNA molecule separate). However, the recent theoretical advances in 
nonlinear theory, related to self-localization in nonlinear networks,
\cite{Scott}
along with the capability to carry out experiments on single 
molecules (to be discussed below) achieved in the last few years 
\cite{Nature} have impulsed a shift on 
the emphasis of the modelling to the dynamics of single molecules.
On the other hand, the decoding of the genome has also led to 
new questions, and, in particular, to the necessity for a better
understanding the relationship between sequence, physical properties,
and biological function.

In this work, we aim to contributing a step in the above described direction
by revisiting the Englander model. Our main goal is to include in the system
the effects of the sequence heterogeneity in order to assess whether the 
model contains the necessary ingredients to reproduce the behavior of 
real DNA molecules. In addition, we want to keep in mind that actual 
experiments (or biological functioning) will unavoidable take place in 
the presence of noise (at least from thermal origin) and therefore another
of our objectives here will be to verify the robustness of the observations 
in the model when noise is considered. In the following sections, we 
review the previous research on which our work is based, then present 
our results for the deterministic case, to close finally with the 
report on the stochastic model. 

\section{Background}

\subsection{Model definition}

\begin{figure}
\begin{center}
\includegraphics[height=4cm]{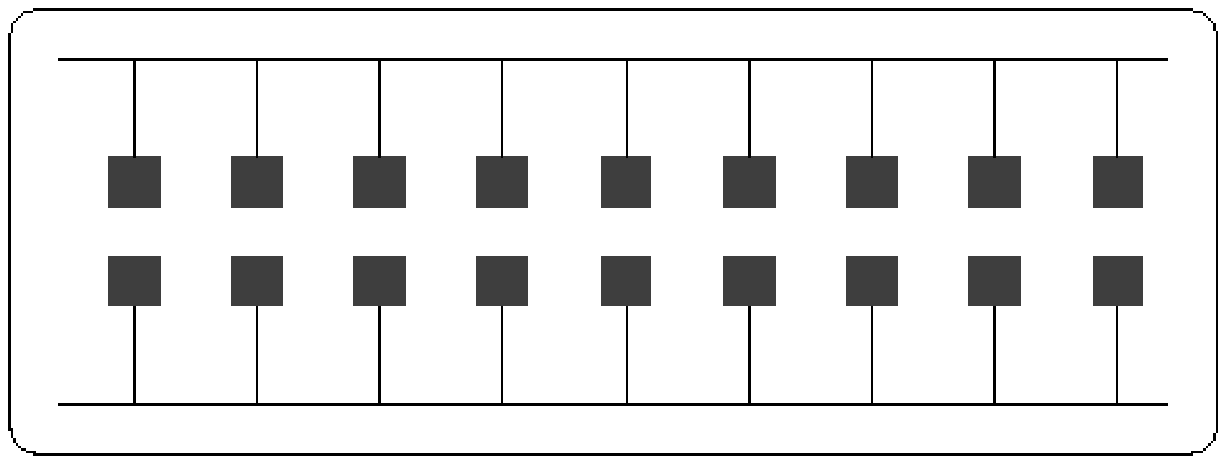}
\includegraphics[height=4cm]{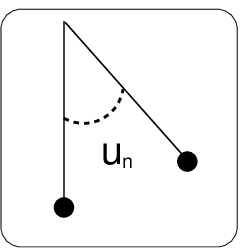}
\caption{\label{fig:modelo}
Sketch of the model. The two strands of the DNA are depicted as 
two straight lines connected by the base pairs (bases are the
squares). The lower strand is assumed to be fixed and as
showed on the right, the angle $u_n$ is the deviation of the
upper base of the $n$-th pair with respect to the lower one. 
}
\end{center}
\end{figure}

\begin{figure}
\begin{center}
\includegraphics[width=10cm,height=4cm]{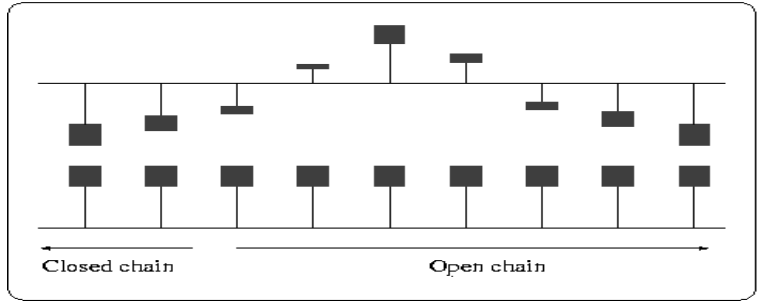}
\caption{\label{fig:kink_modelo}
Kink soliton in the sine-Gordon model. The kink joins a sector of the 
chain where bases are closed $u_n=0$ to another one where bases have
performed a complete turn $u_n=2\pi$. In this last part the chain is 
said to be open if the model is to represent mechanical denaturation.
}
\end{center}
\end{figure}

The model proposed by Englander {\em et al.} \cite{Englander} is 
schematically represented in Fig.\ \ref{fig:modelo}, and is nothing
but a chain of damped pendula, i.e., a discrete, damped sine-Gordon
model, given by  
\begin{equation}
\label{1}
\ddot{u}_n-\frac{1}{a^2}(u_{n+1}-2u_n+u_{n-1})+V_n\sin u_n+\alpha\dot{u}_n+F=0,
\end{equation}
where $a$ is the lattice spacing, $V_n$ is a site dependent constant which 
arises from the specific parameters for the pendulum at site $n$, $\alpha$
is the damping coefficient, and $F$ is a driving term possibly acting on 
the chain. When $V_n=V$ (i.e., we have a homopolymer) and the lattice 
spacing is very small, the system of ordinary differential equations (\ref{1})
can be very well approximated by its continuum limit. Letting
$u_n(t){\longrightarrow} u(x,t)$ when $a^2\ll V$, we find the driven, 
damped sine-Gordon equation:
\begin{equation}
\label{2}
\partial^2_t u-\partial^2_x u+V\sin u+\alpha\partial_t u+F=0.
\end{equation}
It is well known \cite{Scott} that in the absence of dissipation and force 
($\alpha=F=0$), Eq.\ (\ref{2}) possesses soliton solutions 
of the kink type, whose expression is 
\begin{equation}
\label{3}
\phi_{\pm}(x,t)=4\arctan\left\{\exp\left[\pm\sqrt{V}\left(
\frac{x-x_0-vt}{\sqrt{1-v^2}}\right)\right]\right\}
\end{equation}
where the plus or minus sign stands for kinks or antikinks respectively.
If dissipation is included, the soliton is asymptotically stopped, 
whereas the force changes the minima of the onsite potential and 
correspondingly the values of the kink `wings'. If both are present,
as we will see below, the soliton acquires a limit velocity as a
result of the balance of both effects.
Figure \ref{fig:kink_modelo} sketches the physical meaning of a kink 
solution in the context of DNA modelling. 

A comment is in order regarding the physical origin of the force term $F$. 
As stated in the Introduction, the experimental capabilities have been 
improved in the last ten years to the point that it is possible to 
manipulate single DNA molecules with high accuracy. In particular, 
in the so-called mechanical unzipping experiments, one of the DNA strands
is attached to a glass bead, which is pulled by a glass micro-needle,
while the other strand is attached to a glass plate, which serves as 
fixed reference point. The glass bead is then pulled at constant velocity,
opening or unzipping the double chain by consecutively breaking the 
hydrogen bonds of the base pairs, and the force used to pull it is recorded.
The first experiments along this line were carried out in 1997 by 
Essevaz-Roulet and cowokers \cite{essevaz} and largely improved in 
accuracy in 2002 \cite{essevaz2}. Other experiments have been carried out;
the interested reader may find a review by Bustamante, leader of one of the
most important groups in the field, in \cite {Nature}.
The use of a model as ours is in 
fact very appropriate since an analysis of the experimental results
shows that such an experiment can not distinguish individual base 
pairs; instead, it observes the breaking of groups of a few tens
of base pairs, which is correctly represented by the displacement
of one kink \cite{reviewmichel}. In this context, the force $F$ in Eq.\ (\ref{2}) represents
the action of the pulling on the end of one strand on the region 
where unzipping is taking place. We will see in the Conclusion that
our model indeed reproduces qualitatively the experiments.

\subsection{Previous results}

Following the pioneering report of Ref.\ \cite{Englander},
an important extension and application was proposed and studied 
by Salerno \cite{Salerno}. He was the first one to consider the 
effects of the sequence by studying spontaneously travelling kinks 
in T7A$_1$ promoter (a region of the genome preceding a gene where
the transcription activity starts) of the bacteriophage T7. Salerno
studied further other regions of the bacteriophage \cite{Salerno2},
finding that the dynamical activity of kinks should be very special
in the promoters. Recently, the problem was revisited in Ref.\ 
\cite{copia} following the sequencing of the whole genome of the
bacteriophage T7. The main conclusion of this work was that there 
indeed is a significantly higher degree of activity in promoter 
regions, even in the presence of noise. 

The results of the simulations reported in 
Refs.\ \cite{Salerno,Salerno2} were analyzed theoretically by 
Salerno and Kivshar \cite{Salkiv}, who, using a collective
coordinate approach (see, e.g., Ref.\ \cite{SB} for a review on collective 
coordinate techniques for soliton-bearing equations),
developed a description of the
kink dynamics in terms of an effective potential. We will not review in 
detail the procedure to obtain this effective potential but, instead, 
we will simply outline the main steps. The idea is to consider the
undamped model, which can be derived from the Hamiltonian
\begin{equation}
\label{4}
H^{sG}[\{u\}]=\sum_{n=1}^N\{\frac{1}{2}\dot{u}_n^2+
\frac{1}{2a^2}[u_{n+1}-u_n]^2
+V_n(1-\cos u_n)+Fu_n\}. 
\end{equation}
We now insert the following {\em Ansatz}
\begin{equation}
\label{ansatz}
\phi_n(X(t))=4\arctan\left\{\exp\left[\sqrt{V_{avg}}\left(
na-X(t)\right)\right]\right\},
\end{equation}
where $V_{avg}$ is the average value of $V_n$ over the chain, 
in the Hamiltonian (\ref{4}). 
We thus arrive at 
\begin{equation}
\label{5}
E_{eff}=\frac{1}{2}\dot{X}^2+V_{eff}(n, \{V_n\}),
\end{equation}
which is formally equivalent to the energy of a particle subjected to 
the action of the effective potential $V_n$. The exact formula for 
this potential is quite cumbersome and we do not need to reproduce 
it here. We refer the reader interested in the details of the calculation
and the full result to Refs.\ \cite{Salkiv,yo,copia2}.

\subsection{Aperiodic chains}

In this situation, in 1995, when the genome was not yet available, 
F.\ Dom\'\i nguez-Adame and coworkers \cite{yo} propose to mimic 
the behavior of DNA heteropolymers of biological relevance by 
replacing the sequence dependent values $V_n$ by an aperiodic, 
but fully deterministic sequence. As examples, they considered 
mostly the Fibonacci sequence, that is generated according to the
rules $A\to AB$, $B\to A$ repeteadly applied to the initial seed $A$. 
The results obtained in that work (subsequently confirmed in 
Ref.\ \cite{copia2}) pointed out the existence of an intriguing 
phenomenon, namely the existence of a threshold force for a kink 
to start moving along the chain. Once the force is above the threshold,
the kink moves with an asymptotically constant given by the balance 
of damping and driving. An analytical expression can be easily derived
to predict the asymptotic velocity from energy conservation arguments
(see Ref.\ \cite{McL}). Although the original argument is valid for a
homogeneous chain, repeating it here with the average value $V_{avg}$ 
leads to a prediction which is very accurate for periodically ordered
chains while overestimating (by approximately a 10\%) 
the velocity for the Fibonacci chains.
\begin{figure}
\begin{center}
\includegraphics[height=7cm]{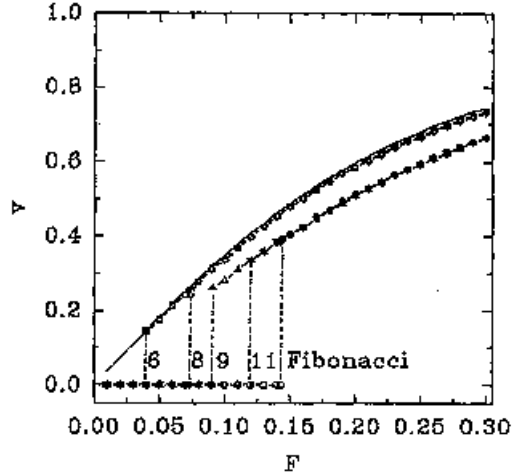}
\caption{\label{fig:Fibo}
Steady state velocity versus applied force for chains with different orderings.
Points correspond to numerical results, numbers indicate the Fibonacci 
generation constituting the unit cell of the chain. Reprinted from 
Ref.\ \protect\cite{yo}; see the original paper for more details.
}
\end{center}
\end{figure}
Even more interestingly, it was also found (see Fig.\ \ref{fig:Fibo}) that 
the threshold value depended on the length of the quasiperiodic chain 
considered. This was seen by forming periodic chains with unit cell 
$F_n$, $F_n$ being the Fibonacci chain in iteration $n$. As shown in 
Fig.\ \ref{fig:Fibo}, increasing the size of the unit cell leads to 
an increasing of the threshold, reaching a limiting value for
the whole Fibonacci chain. This behavior was well described with
the help of the effective potential approach summarized above.

Based on this results and on the fact that, for the same parameters,
random chains exhibited much larger threshold values (around $F\simeq 0.5$),
Dom\'\i nguez-Adame {\em et al.} conclude that long range order effects
give rise to measurable consequences on kink dynamics in aperiodic chains.
Further, they suggested that the fact that there are long range correlations
in DNA and, in any event, that it contains information and is not purely 
random, could lead to similar phenomenology in the propagation of nonlinear
coherent excitations along the molecule. At this point is where we take 
that previous research, with the body of genomic data available nowadays. 

\section{Effects of the sequence}

In order to check whether or not the phenomenology we have summarized in 
the previous section carries over to real DNA data, we have simulated 
Eq.\ (\ref{1}) with $V_n$ chosen according to sequences obtained from 
the human genome. We want to stress that this description of DNA,  
arising from the original model of Englander and coworkers \cite{Englander},
intends to be only qualitatively correct. Therefore, the parameters can 
be freely chosen, trying, of course, to mimic the real ones. Therefore,
what we do is choose $V_n=2,3$ according to what we have at site $n$ is
an A-T pair, linked by two hydrogen bonds, or a C-G pair, with three 
hydrogen bonds. The other parameters are chosen as in Ref.\ \cite{yo},
namely lattice spacing $a=0.1$ and damping $\alpha=0.1$. With respect to 
the lattice spacing, we want to point out that the chosen value leads to
a width of the kink which is comparable to that of spontaneous openings
of real DNA chains. In what follows, we use data obtained from the
National Center for Biotechnology Information for the human genome 
({\tt http://www.ncbi.nlm.nih.gov}).

\begin{figure}
\begin{center}
\includegraphics[height=7cm,angle=270]{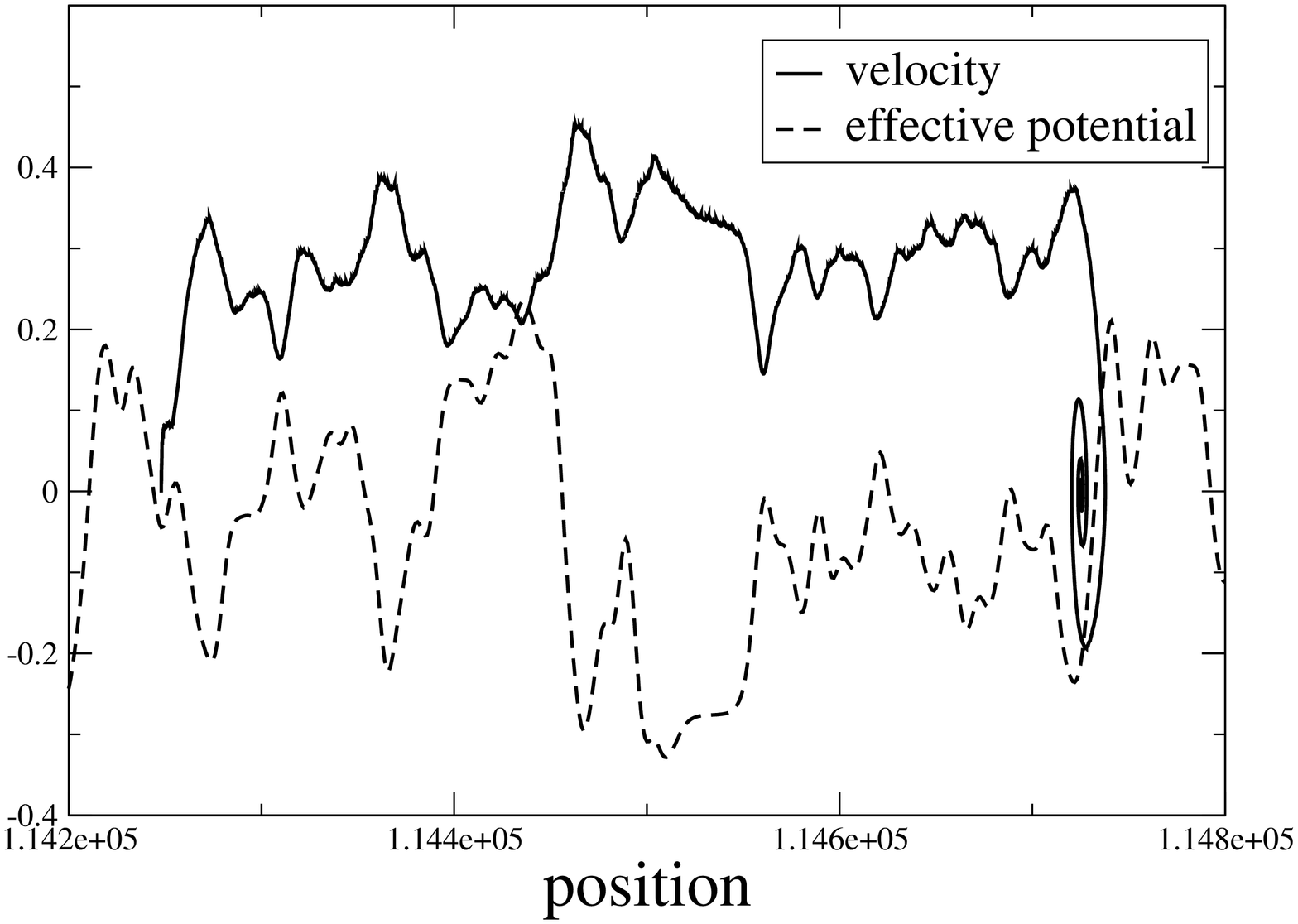}
\includegraphics[height=7cm,angle=270]{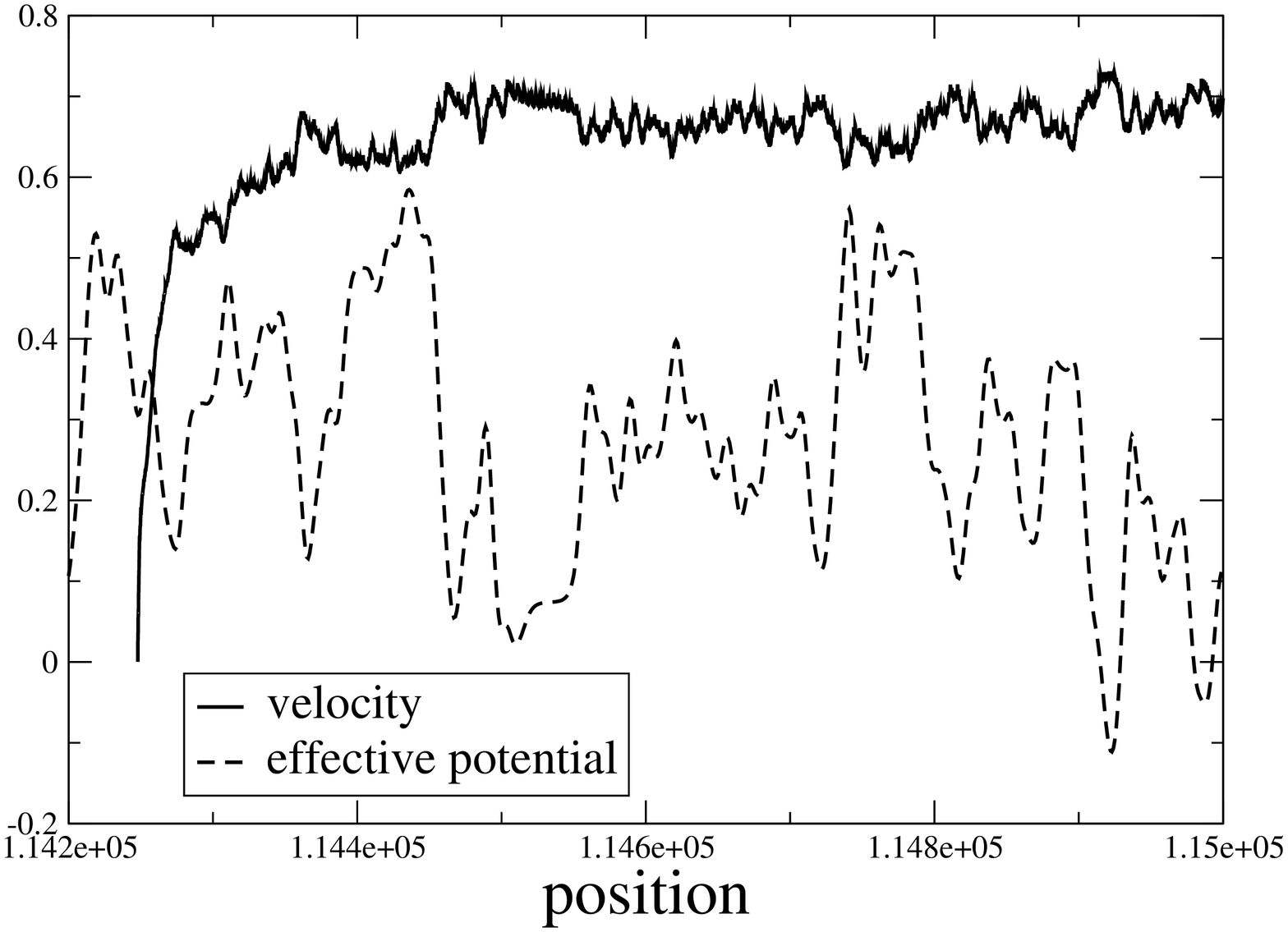}
\caption{\label{fig:velo}
Simulations of the kink soliton in the sine-Gordon model with genome data.
Shown are the velocity and the (properly scaled to fit in the plot) 
effective potential vs the position along the chain. Left: $F=0.06$;
the kink travels along the sequence but stops at an effective potential
well.  Right: $F=0.07$; the kink travels along the whole chain.
The DNA sequence corresponds to 
contig NT\_028395.2 of human cromosome 22, between positions 114\,100 
and 115\,100, part of a gen.
}
\end{center}
\end{figure}

We have carried out simulations on many different sequences of different 
chromosomes, choosing both coding and non-coding regions.
Typical results from our simulations are plotted in Figs.\ \ref{fig:velo} 
and \ref{fig:for}; the outcome from all the regions analyzed is always
qualitatively the same. Figure \ref{fig:velo}
shows the kink dynamics on a real dna sequence when the kink is forced 
with two different drivings. For reference, we plot in the same graphs
the effective potential obtained as described in the previous section. 
It is important to note that the effective potential does not contain 
the contribution of the force $F$, and therefore its most relevant 
information is the position of the peaks and valleys. With this in 
mind, we see that the dynamics is basically as in the case of the 
Fibonacci chain. There is a threshold for the kink to propagate along
the whole chain and, for forces below threshold, it ends up being 
trapped at some potential well. 

Figure \ref{fig:for} collects our observations regarding the 
existence of thresholds in two examples:
\begin{figure}
\begin{center}
\includegraphics[height=7cm,angle=270]{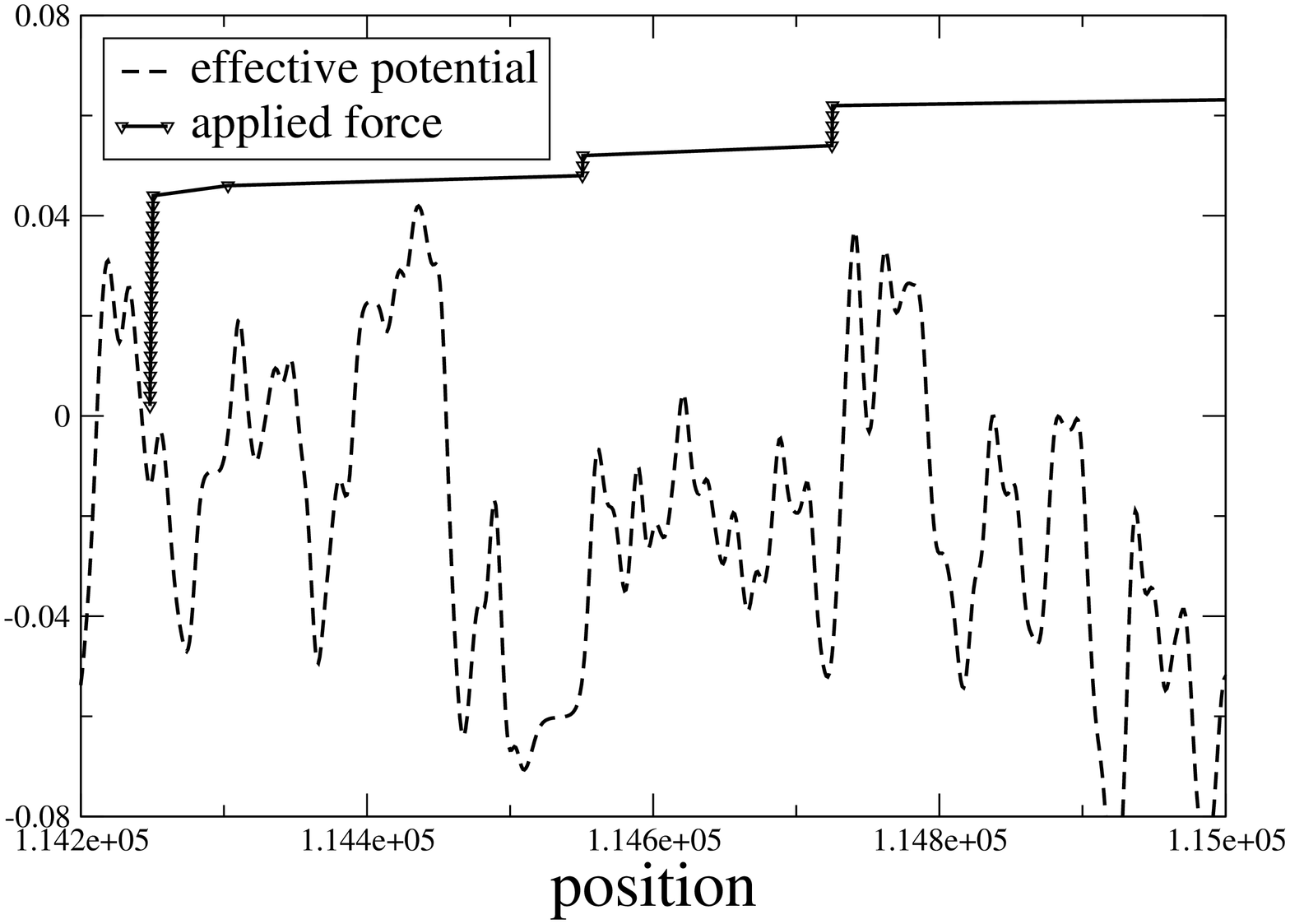}
\includegraphics[height=7cm,angle=270]{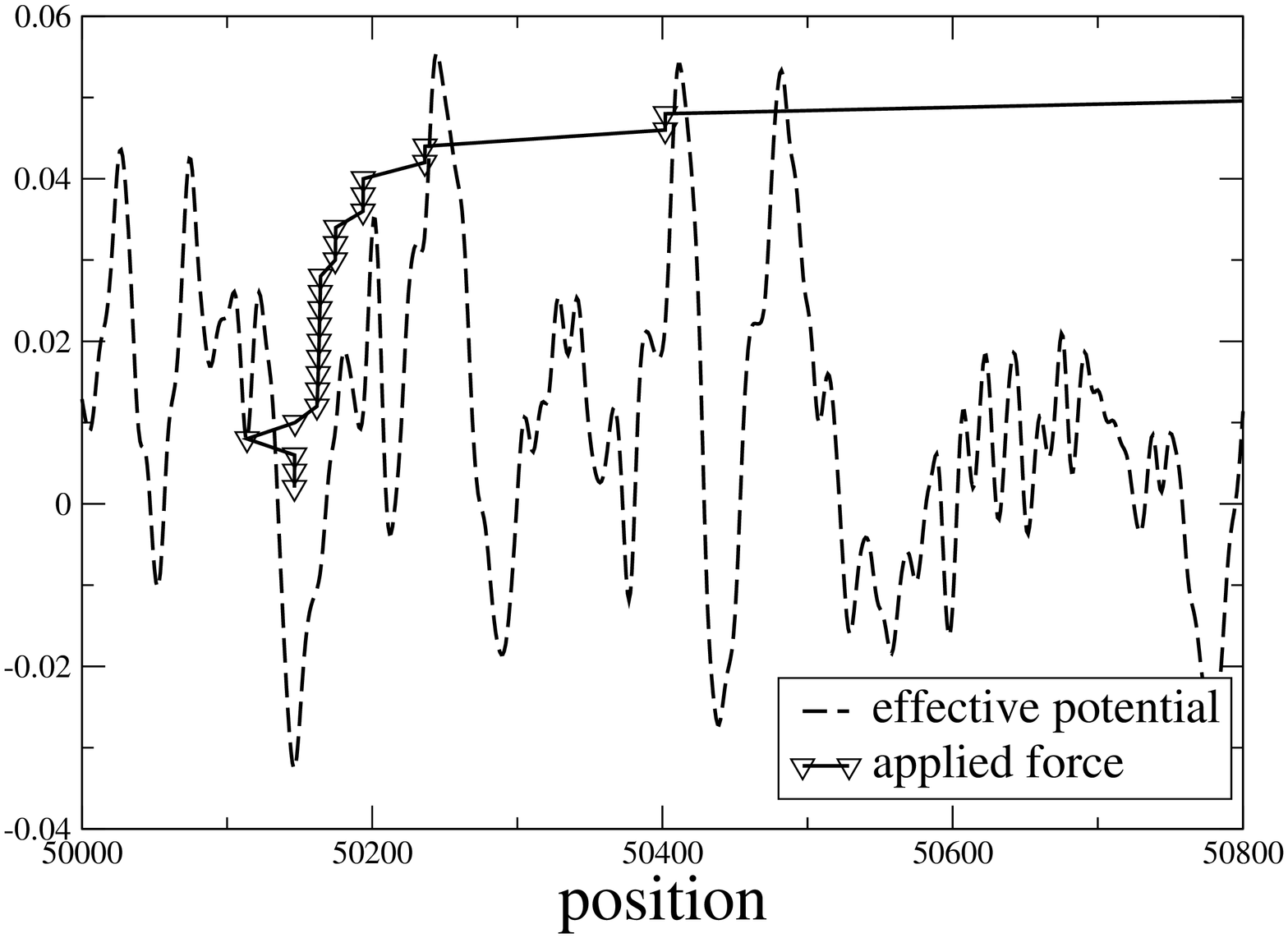}
\caption{\label{fig:for}
Simulations of the kink soliton in the sine-Gordon model with genome data.
Shown are the force required to reach a certain position and the 
(properly scaled to fit in the plot) effective potential vs the 
position along the chain. Thresholds correspond to the largest force 
for which there are points in the plot. 
Comparison of propagation along a coding region (left, same contig as in Fig.\ 
\ref{fig:velo}, positions 114\,000 to 115\,000) and a non-coding region
(right, same contig as in Fig.\
\ref{fig:velo}, positions 50\,000 to 51\,000).
}
\end{center}
\end{figure}
a coding region and a non-coding region. We again see the existence of 
threshold forces, in agreement with a description in terms of an 
effective potential. However, we do not observe any qualitative or 
otherwise relevant difference between the kink dynamics in the two 
regions. This is the case with all the regions we have analyzed. 
Therefore, the hint from the Fibonacci results that information may 
lead to different kink dynamical properties is, at least at the level 
of our simple model, not in agreement with the simulation results. 

\begin{figure}
\begin{center}
\includegraphics[height=10cm,angle=270]{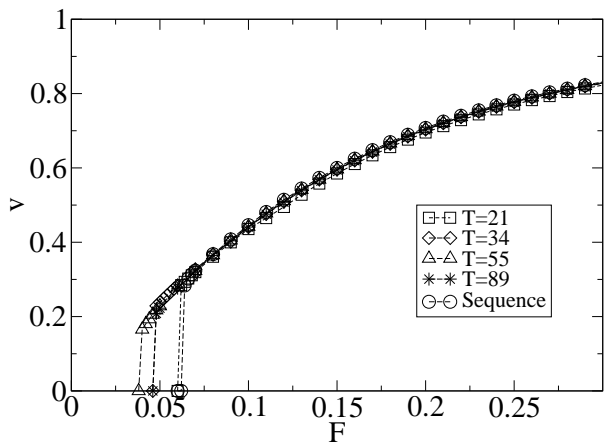}
\caption{\label{fig:seq}
Velocity vs applied force for different periodic systems 
formed by repeating as a unit cell $n$ sites of the 
same contig as in Fig.\ \ref{fig:velo} beginning at 
site 114\,241, with $n=21$, 34, 55 and 89 (as in the 
Fibonacci case) and for the whole sequence going from
114\,100 to 115\,100.
}
\end{center}
\end{figure}

In order to assess further how good is the comparison between 
the previous work on the Fibonacci chain and the present results,
we have simulated periodic systems with unit cell built from 
pieces of a genetic sequence, repeated to complete a longer 
chain. The sizes for the unit cells were chosen to mimic the 
sizes for the Fibonacci iterations. The results are collected 
in Fig.\ \ref{fig:seq}.
{}From this plot, we see that the most important difference between
the two cases is that in the actual DNA chain, the threshold value
does not depend monotonously on the size of the unit cell (note that
the threshold for the periodic chain with unit cell of size 55 is the
smallest one). It is still true, however, that the threshold is largest for the
whole sequence. Therefore, we are led to conclude that the information 
effect on the kink propagation along DNA chains is certainly different 
of what we expected in terms of our simple Fibonacci approach. 
Notwithstanding, we want to stress that the analytical approach in 
terms of collective coordinates and an effective potential is a 
very good picture of the observed phenomenology even in this, more 
realistic setting. On the basis of this scenario, we suggest 
that the reason for the difference between the two models lies
in the larger diversity 
available for possible DNA chains. The fact that sites of type 
$B$ in Fibonacci chains are always isolated is very restrictive 
as to the shapes and sizes of the peaks of wells of the potential
felt by the kinks, and therefore the Fibonacci model cannot capture
the richness of the DNA model. As for the possible different behaviors
between coding and non-coding regions, we conclude that the Englander 
model does not reflect the information content of the chain in the
kink dynamics, which in principle prevents the use of this simple 
model as a genome sequencing tool. 

\section{Effects of noise}

The results discussed so far correspond to the purely 
deterministic case, i.e., they are related only to the 
``disorder'' (understood as inhomogeneity, even if it 
contains information) present in the DNA chains. However, 
a fully deterministic approach is not really relevant to 
the DNA properties in so far as life takes place at 
nonzero temperature, and the same applies to single 
molecule experiments, usually carried out at room temperature. 
It goes without saying that there are many aspects of the 
influence of noise on the dynamics of the nonlinear excitations
in DNA, and that our purpose is simply to present the first 
results of two lines of research we are beginning to explore 
in this context. Below we discuss the robustness of the deterministic results 
presented in the preceding section when thermal noise is considered, 
and how nonlinear coherent excitations arise from unstructured, noisy
distortions of the DNA chain. 

\subsection{Thermal noise}

Our first aim is to check the validity of the general description
of the kink dynamics in terms of the effective potential in the 
presence of thermal noise. The reason for this is that, unless the
scenario holds at temperatures of the order of room temperature,
the insights provided by the effective potential about special 
places in the chain will not be relevant for the actual behavior
of DNA, either {\em in vivo} or in single molecule experiments. 
To this end, we considered the Langevin version of Eq.\ (\ref{2}),
namely
\begin{equation}
\label{2bis}
\partial^2_t u-\partial^2_x u+V\sin u+\alpha\partial_t u+F+\xi(x,t)=0,
\end{equation}
where $\xi(x,t)$ is a gaussian white noise, of zero mean and 
variance 
\begin{equation}
\label{ruido}
\langle\xi(x,t)\xi(x',t')\rangle=2D\alpha\delta(x-x')\delta(t-t'),
\end{equation}
verifying the fluctuation-dissipation theorem. 

\begin{figure}
\begin{center}
\includegraphics[height=7cm]{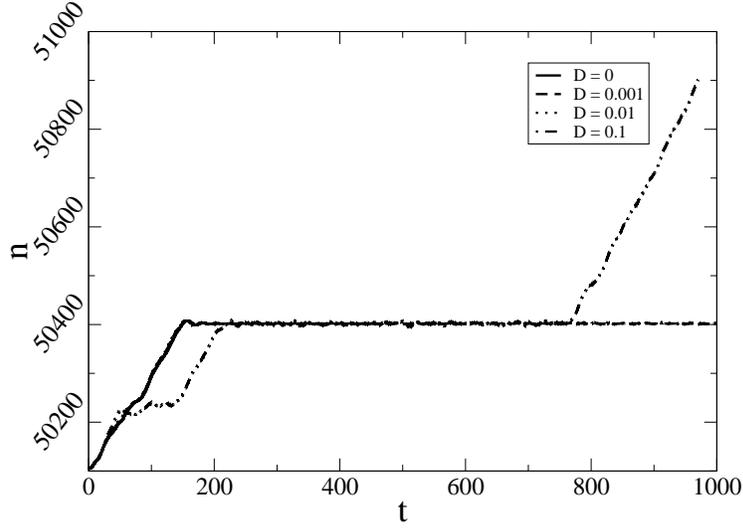}
\caption{\label{fig:noise}
Kink soliton propagation in the presence of noise. The sequence is
as in Fig.\ \ref{fig:velo}. The applied force is $F=0.048$ 
(deterministic threshold for this sequence is $F=0.05$). 
Noise intensities $D$ are as indicated in the plot. 
}
\end{center}
\end{figure}

The results obtained in the presence of noise are illustrated by 
the example in Fig.\ \ref{fig:noise}.
The plot represents the trajectory of the center of a kink for 
different values of the noise intensity $D$. The applied force 
$F$ is below threshold, so for $D=0$ we observe that the kink is
trapped at a position around the base pair 50400. As we may see,
it is necessary to add a large amount of noise (note that the 
value of $D$ is twice the value of the threshold force) to 
allow the kink to escape that well, and even then this takes place
after a rather long residence time. Other runs yield similar 
results. Therefore, our preliminary conclusion is that the 
effective potential information is relevant even in the presence
of moderate or large noises. Pending a more detailed comparison 
with {\em in vivo} temperatures, we feel that the available data
is enough to ascertain that the model will behave qualitatively 
as single molecule, mechanical denaturation experiments carried 
out at room temperature. 

\subsection{Formation of nonlinear excitations and 
transcription initiation}

In this subsection, we deal with the problem of the formation of
nonlinear excitations in the model. This is an important issue in
view of the fact that we are studying kink propagation using 
a kink as an initial condition, but we have not justified whether
or not these kinks can build up from a generic excitation of the
chain. For this study, 
we have considered one of the sequences studied in 
Ref.\ \cite{LosAlamos}. Specifically, we have taken the sequence
of adenoassociated virus 1 (AAV1), consisting of 4718 base pairs. 
The sequence contains the adenoassociated viral P5 promoter studied
in Ref.\ \cite{LosAlamos}. Kalosakas {\em et al.} performed 
Langevin simulations of a very short chain containing only the 
promoter and found that the chain exhibited preferential opening 
precisely at the so-called transcription start site. From their 
data, they concluded that DNA dynamically directs its own 
transcription. By looking a the same sequence data these authors 
have analyzed, we simultaneously address the issue of the origin 
of nonlinear excitations and the confirmation of the results
of Ref.\ \cite{LosAlamos}. 

In our simulations, we have again considered the deterministic 
equation (\ref{2}), which we now integrate starting with 
random initial data given by $u_n=u_0+v_n$, where $v_n$ 
are independently generated gaussian random numbers. 
We have set the applied force $F$ to zero in order to check 
whether breathers and kink can be formed in the absence of 
directed external pulling, only from randomly distributed 
energy. Dissipation is also zero in these simulations because
otherwise all breathers would be suppressed in the long term
(while they survive in the presence of thermal noise, 
see \cite{LosAlamos}).
We have taken free boundary conditions, which implies that 
energy is not conserved. The reason for this choice is twofold:
On the one hand, if periodic boundary conditions are used, a large
amount of radiation is all the time present in the system, forming
a background that masks the static breathers, formed and pinned 
by the sequence (this is not the case with kinks of course, but
in many simulations we obtain only breathers). Free boundary 
conditions let this radiation go away from the system allowing to
identify the relevant pinned breathers. On the other hand, in 
experiments with single DNA molecules the energy in a single 
molecule is not conserved, as the chain deformations may be 
resolved through the free ends of the molecules. Therefore, by
using free boundary conditions we are mimicking more closely 
the experimental situation. We note that in doing so we are not 
in the microcanonical ensemble of statistical mechanics; this is 
not a problem in so far as we are only interested in statistics 
as a way to quantify how often breathers or kinks form at specific
places. 
Finally, we mention that our chain consists of 
the base pairs from position 200 to 400 (the P5 promoter goes from
position 256 to 324, and the transcription start site
is at position 296. 

\begin{figure}
\begin{center}
\includegraphics[height=8cm]{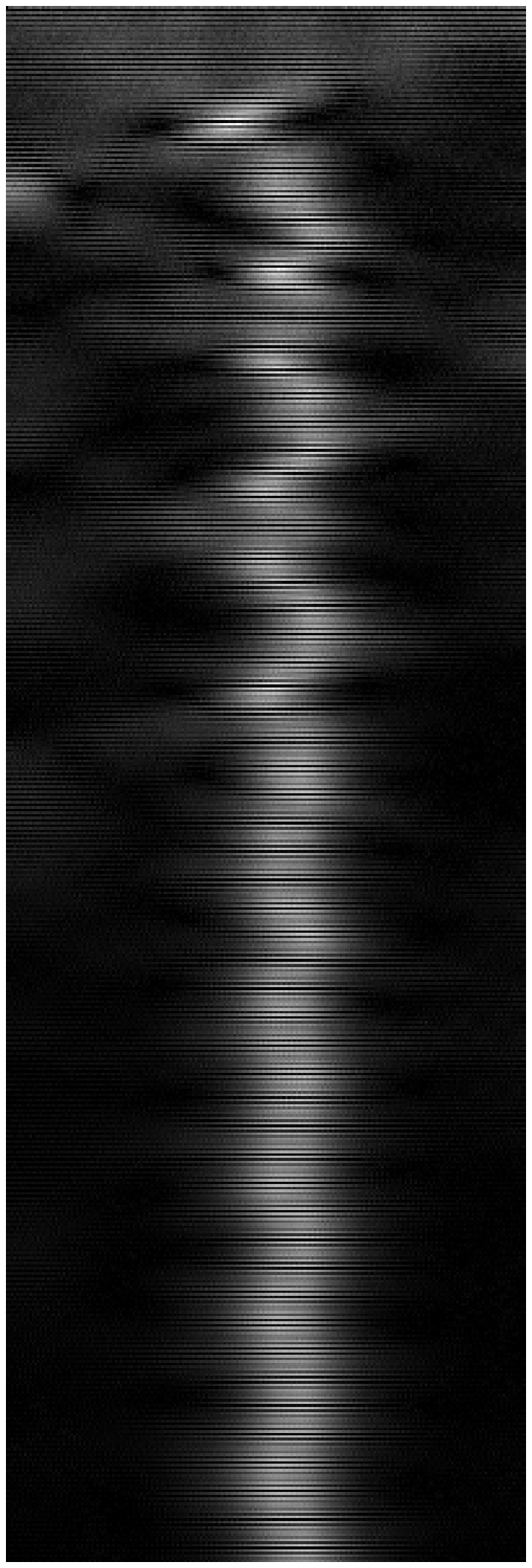}
\includegraphics[height=8cm]{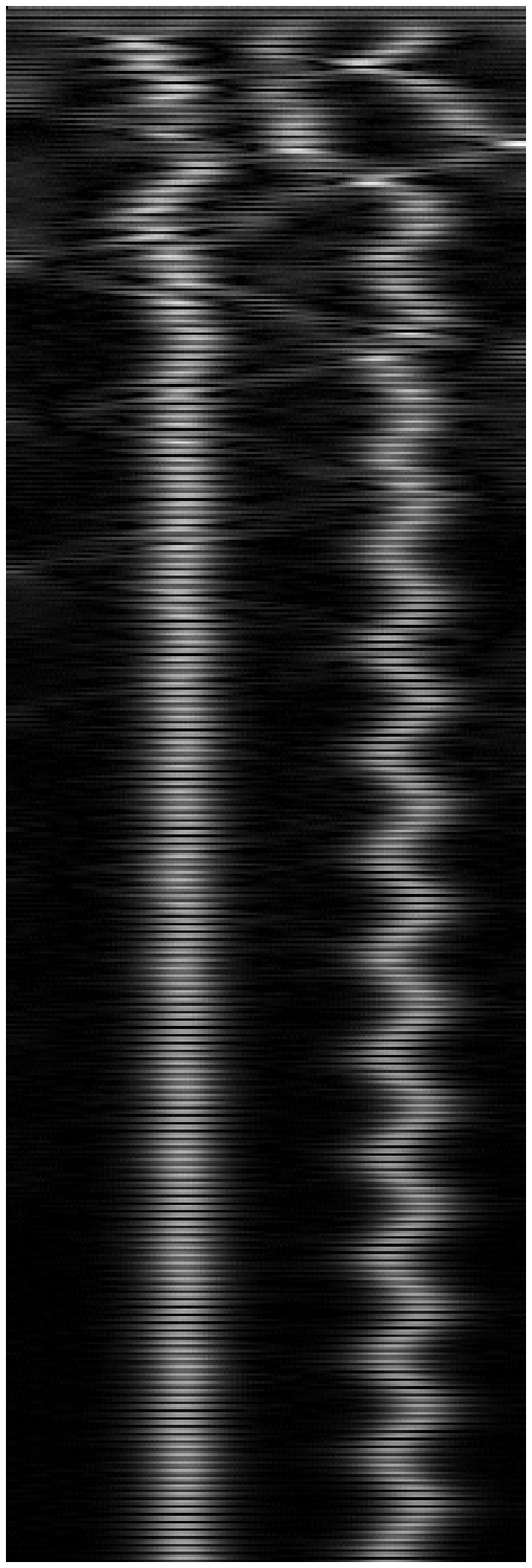}
\includegraphics[height=8cm]{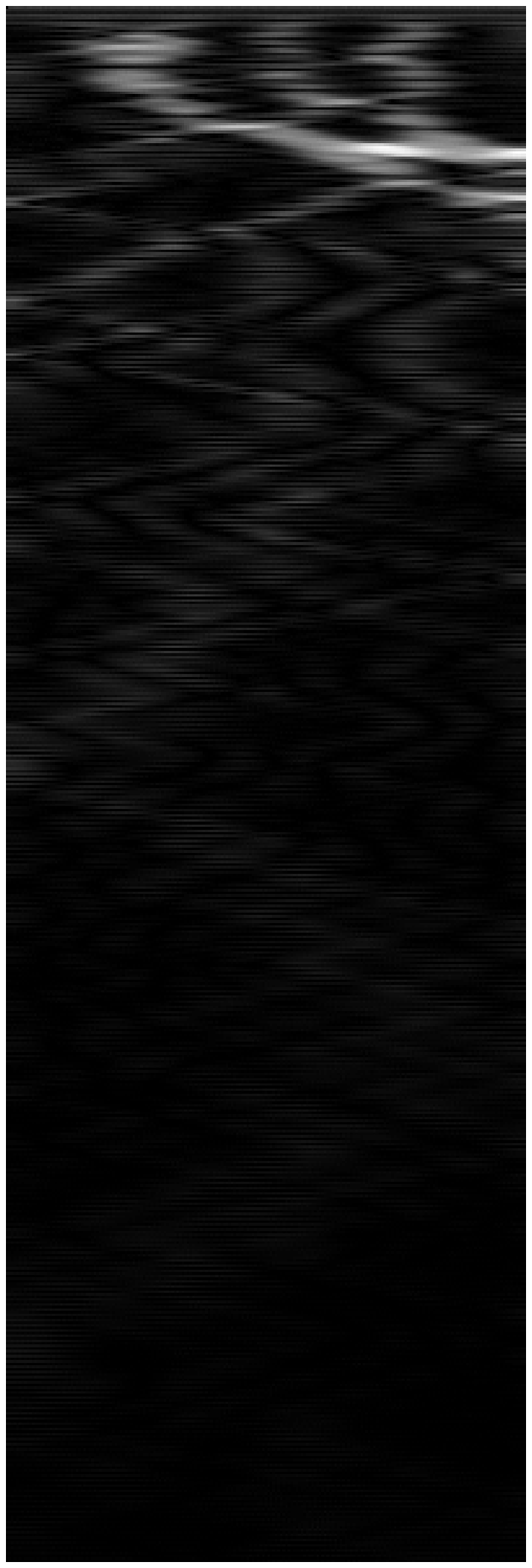}
\includegraphics[height=8cm]{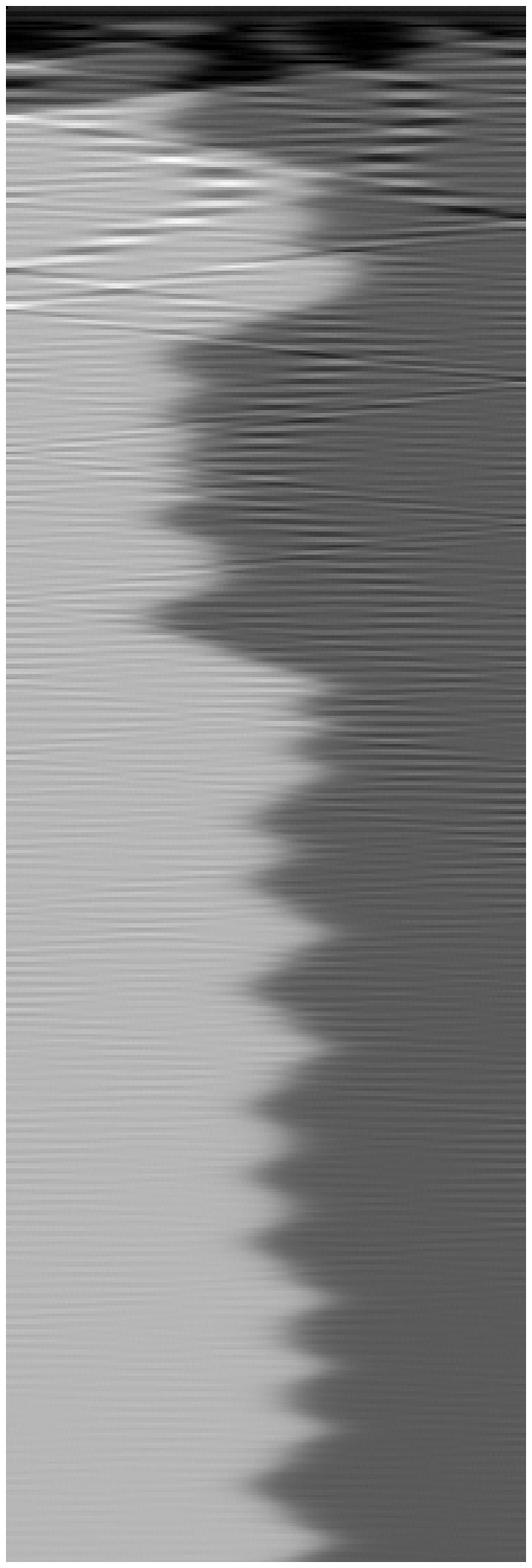}
\caption{\label{fig:grey}
Focusing of random initial data. Grey levels code the absolute value of 
$u$ along the chain horizontally, and time proceeds downwards. Left to 
right, the initial data is $u_0=0.9$, $2.0$, $2.5$, and $3.0$. 
Superimposed noise variance is 0.1. Damping is set to zero. 
Sequence corresponds to AAV1 from position 200 to 400 (see text). 
}
\end{center}
\end{figure}

Figure \ref{fig:grey} presents an example of the time evolution 
of random initial data.
Although the plots correspond to a specific set of parameters, the 
different outcomes represent those of typical simulations. The case
represented here illustrates the effect of changing the mean value of
the initial data, $u_0$. In general, what we observe is that the initial data 
may or may not give rise to localized, coherent excitations, and that these
can be breather-like or kinks (these last ones appear only if the value 
of $u_0$ is large, typically close to $\pi$). Our statistics so far is not
large enough to give reliable values for the relative frequencies of each
event, but we have seen that none of these outcomes is rare, i.e., all of 
them are statistically relevant. Hence, these simulations demonstrate that
it is indeed possible to have nonlinear coherent excitations in DNA chains,
formed spontaneously from unstructured initial data. 

As regards the comparison with Ref.\ \cite{LosAlamos}, we have found that
when breathers are formed, they preferently locate in the transcription 
start site or its neighborhood. Actually, in all the simulations we have 
carried out, we observe that breathers only build up in three different 
regions of the studied sequence. This agrees with the findings reported in
Ref.\ \cite{LosAlamos}, although we need to do a much larger statistical
analysis. On the other hand, preliminary analytical results show that the
collective coordinate approach leading to the effective potential for kinks
can be (approximately) extended to breathers. This extension, which will be
reported elsewhere, predicts with good accuracy the existence of three 
potential wells roughly at the places where we observe the formation of 
breathers in the simulations. If this is indeed the case, the process would
develop in two stages: first, breathers are formed by focusing of the initial
data, and second, these breathers experience the effective potential and 
end up located at one of its wells. While this is an appealing picture, 
much more analytical work is needed to confirm it, and this preliminary 
report should still be taken with some degree of caution. 

\section{Conclusions}

In this paper, we have reported the first conclusions of our work on the 
effects of the genetic sequence on the propagation of nonlinear excitations
along (simple models of) DNA chains. The research was motivated by previous 
results on chains given by aperiodic sequences \cite{yo}, which lead to 
the hypothesis that the information content of the inhomogeneities might 
be relevant for the dynamical behavior. While this general statement cannot
be ruled out at this stage, we believe that at the level of the Englander 
model \cite{Englander} the simulations presented here do not confirm that
hypothesis. Specifically, we have not observed qualitative difference in
the dynamics of kinks propagating along coding or non-coding regions. 
In this respect, it is important to realize that the function and information
content of non-coding regions are far from well know, and it may well be 
that the dynamics in the two types of sequences is the same because the
quantity of information is similar. This is certainly an interesting issue
that deserves further attention. On the positive side, we have seen that
a model as simple as ours is able to reproduce the main stylized features
observed in mechanical unzipping experiments
\cite{reviewmichel,Nature,essevaz,essevaz2},
such as the existence of a threshold to begin the unzipping process and 
the irregular time behavior of the force, which closely resembles stick-slip
dynamics. This agreement suggests that, in modelling DNA, the most relevant
factor is the correct inclusion of the `disorder' given by the sequence
rather than the mechano-chemical details of the interactions.

Another conclusion that we deduce from this research is the validity of the
effective potential approach\cite{Salkiv,yo,copia2}
to the dynamics of nonlinear excitations in DNA. Although the precise 
dynamics of kinks is difficult to predict, this analytical approximation 
allows to identify the possible stop positions as well as the barriers 
that control the total opening (mechanical denaturation) of the chain. 
Furthermore, the scenario that arises from considering the effective 
potential a good description of the kink dynamics leads us to speculate
that the results of mechanical denaturation experiments on single 
molecules, with their stick-slip-like time evolution, could be also 
understood within simple models like the one studied here. This 
impression is reinforced by the fact that the effective potential 
approach works equally well in the presence of (thermal) noise, and
by our preliminary report of its validity for breathers. Work along
these lines is in progress. 

Finally, a few words are in order regarding our findings about the 
preferential opening of DNA chains. We have qualitatively confirmed
the results reported in Ref.\ \cite{LosAlamos} with a much more 
elaborated and realistic model \cite{PB,DPB,DP}. We want to stress 
that our study is not a Langevin simulation and is not even a 
microcanonical simulation because of the free boundary conditions. 
With those caveats in mind, we can conclude that in the sequence 
we have studied there are three preferential places for breather
(opening) formation, one of the most visited ones being the neighborhood
of the transcription start site. In this respect, we want to point out
that those locations may also be predicted by the effective potential 
approach. If that is indeed the case, it could well be that the analyses
of longer chains would also reveal other preferential places whose 
significance should then be understood. In any event, we stress that
our conclusions on this subject are very preliminary and that we 
believe that much more research is needed before we can reach 
definitive conclusions.

\section*{Acknowledgments}
 
We thank Michel Peyrard for many discussions about this work and for his 
patient explanations on DNA dynamics. 
We also thank Alan R.\ Bishop for making the results in Ref.\ 
\cite{LosAlamos} available prior to publication.
This work has been
supported by the Ministerio de Ciencia y Tecnolog\'\i a of Spain
through grant BFM2003-07749-C05-01. S.C. is supported by 
a fellowship from the Consejer\'\i a de Educaci\'on de la
Comunidad Aut\'onoma de Madrid and the Fondo Social Europeo.


\end{document}